# Tony Skyrme and the Origins of Skyrmions[†]

Ian J. R. Aitchison

SLAC National Accelerator Laboratory and Oxford University

## Abstract

I first discuss the main motivations for Tony Skyrme's highly original program (1958-62) of making fermionic nucleons out of bosonic pion fields, as described in his Cosener's House talk in 1984. These include a dislike of point-like elementary particles, which he blamed for infinite renormalization, and a preference for extended objects distinguished by what we now call conserved topological quantum numbers. In this he was strongly influenced by William Thomson (Lod Kelvin), who was so impressed by Helmholtz's proof of the conservation of circulation ("Wirbelbewegung") in fluid vortices that he developed an entire theory of atoms as knotted vortex rings in the ether fluid. Skyrme liked mechanical models, as did Kelvin, and he grew up fascinated by the ingenuity of Kelvin's machine for predicting tides, an example of which stood in his grandfather's house. This seems to have been connected to his strong preference for bosonic fields, which have a classical limit, over fermionic fields which do not. I then sketch the progress of Skyrme's ideas in the series of sis papers in the years 1958-62, which passed largely unnoticed at the time. I emphasize his remarkable intuition that the kink solution of the classical Sine-Gordon equation would be a fermion when quantized; and the novelty of his identification of the Skyrmion winding number with baryon number in the three-dimensional case. I end by briefly describing how Skyrme's work was dramatically related to QCD in 1983-4.

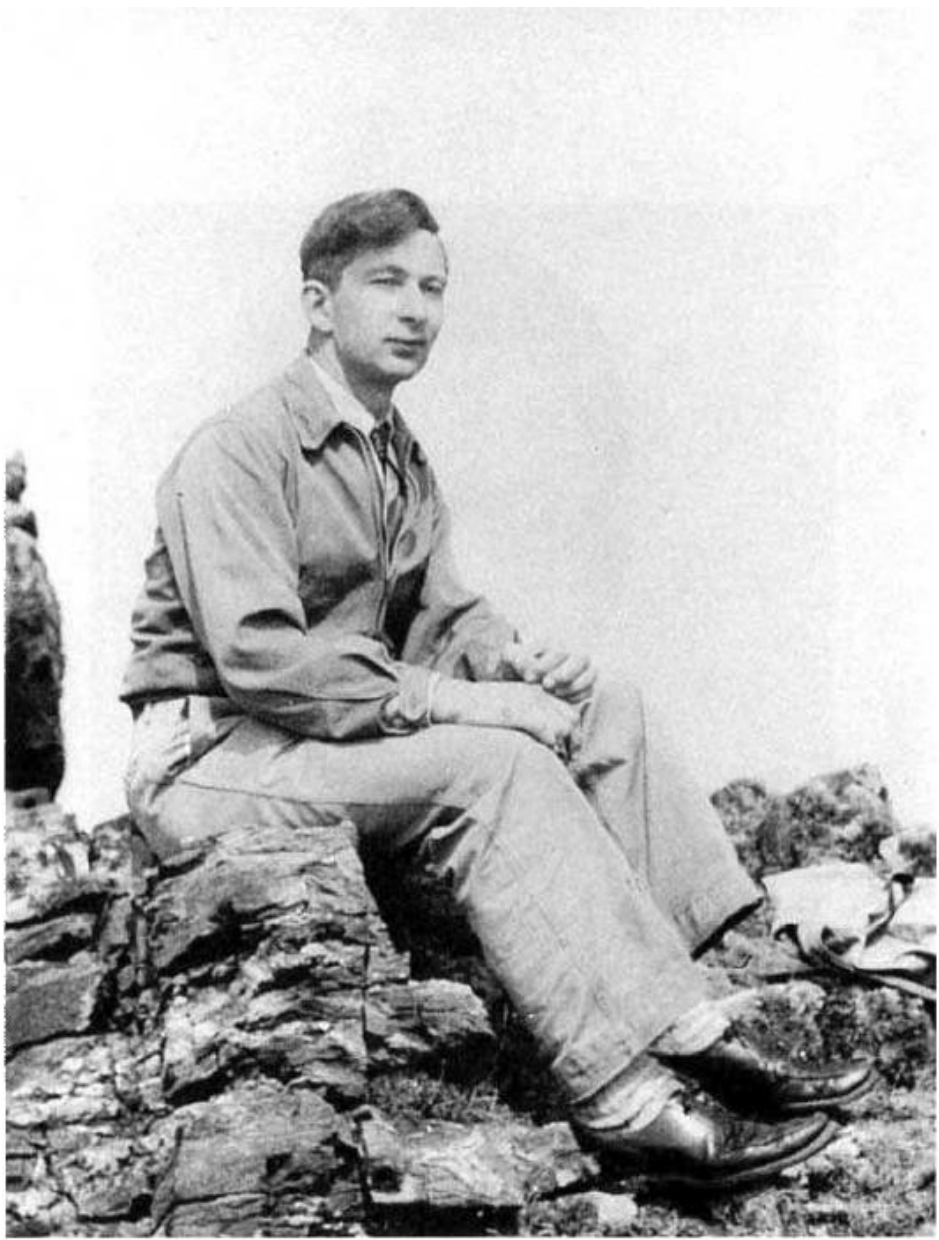

*Taken in 1948 at Cornwall by Dr. Dorothy Skyrme.*

I'd like to begin by briefly explaining how I come to be talking to you about Tony Skyrme and the origins of Skyrmions. Tony's six papers [1-6] on what are now called Skyrmions were published in the years 1958-1962, but attracted relatively little attention for over twenty years. Then, in 1983, Ed Witten [7,8] showed how Skyrme's non-linear meson field theory could be understood as an approximation to the known theory of strong interactions, QCD, with Skyrme's mesonic solitons playing the role of baryons. Like many others, I was captivated by the originality of this idea. Together with Caroline Fraser, then a Junior Research Fellow at Somerville College, Oxford, I began working on aspects of Skyrmions in particle physics. It was a very rapidly moving field, and in November 1984 we helped organize a workshop on Skyrmions at Cosener's House in Abingdon, UK. Naturally we wanted Tony to give a talk, but he was famously reclusive, and reluctant to publish or to give seminars. But we did persuade him to come, and he gave a fascinating talk, in which he essentially let his audience into his private motivations for what was, surely, an extraordinary project – making fermionic nucleons out of bosonic meson fields! I will return to his Cosener's House talk in just a moment.

Tony's work on Skyrmions was done at Britain's Atomic Energy Research Establishment (AERE) Harwell, where he had a position from 1950 to 1961. In 1961 he left Harwell and took up a post at the University of Malaya in Kuala Lumpur. Then, in 1963, Rudolf Peierls' chair of Mathematical Physics at Birmingham fell vacant, when Peierls moved to Oxford, and Tony was appointed as his successor. So in 1987, when Peierls' 80$^{th}$ birthday was celebrated with a Symposium at Oxford, it was natural to hope that Tony could be prevailed upon to give a talk at the Symposium along the same lines as his Cosener's House talk – and this he agreed to do. But two days before the start of the Symposium, the organizers learned that Tony had died, after complications following an unexpected operation. Partly as a memorial to Tony, it was decided to try and reconstruct his Cosener's House talk, using some notes found in his office, and the notes of myself and others who had heard him speak. It fell to me to put this reconstruction together, which was then included in the Peierls Symposium Proceedings [9]. These Proceedings include a brief biography of Skyrme by Dick Dalitz [10], who also provided an outline of Skyrme's life and work [11].

My talk will fall into three parts. In the first part, I'll describe what Tony told us at the Cosener's House meeting in 1984, amplifying some of the historical details. The focus will be on Skyrme's motivations for undertaking what was a highly unconventional, even perhaps quixotic, program. In the second part of my talk, I'll briefly review the series of six papers published in 1958-62, referred to as I – VI. I'll emphasize how he implemented his program, being guided by those motivations. I'll end by very briefly outlining what happened to this program after 1984.

So now I turn to what Tony told us in 1984. He laid out four main motives for his project of making nucleons out of meson fields. The first of these motives was **unification**. For Tony, this did not mean searching for some grand symmetry group. Rather, it meant a desire to build everything from one kind of stuff. Conventionally, there are two types of fundamental fields, bosons and fermions. It would be nice, he said, if you could manage with only one kind. He alluded to Heisenberg's non-linear spinor field theory of 1958, in which everything would be made from a self-interacting fermion field. But Skyrme didn't really like fermions, a point I'll return to in a moment. So, as he said, he thought it might be fun to see if he could get everything out of a self-interacting boson field theory instead. On the face of it, it is hard to imagine how this might work: you can easily add half-integers to make whole integers, but how do you make half-integers out of whole ones?

The second motive was the aim of **getting rid of infinite renormalization**, a problem he traced to the point-like nature of conventional elementary particles. For Skyrme, the word "particle" would mean an extended object, with finite self-energy. In this he was strongly influenced by ideas of the nineteenth century Scottish theoretical physicist William Thomson, later Lord Kelvin. Kelvin will figure prominently in this story.

A third motive was what Skyrme called the "**fermion problem**". He felt unease about any quantum-mechanical concepts that did not have clear classical analogues. He thought that handling fermions via Grassmann variables in the path integral formalism was an unnatural, purely mathematical construction. He hoped to show that the use of fermionic fields would turn out to be an idealized way of describing a certain semi-classical field configuration.

Finally, although Skyrme did not explicitly list this motive in his Cosener's House talk, he told us how fascinated he had been by Kelvin's vortex theory of atoms, which located the origin of a conserved quantity in a structural property of the object, rather than in a symmetry principle. This preoccupation would ultimately lead to one of Skyrme's major contributions: that of a **topological quantum number**.

A common thread in these motivations is a strong preference for a description in terms of fields that have a classical limit – one might almost say a desire for a "mechanical" picture. This was precisely the position of someone whom Skyrme emphasized had a major influence on him: the Scottish mathematical physicist William Thomson, later Lord Kelvin. Thomson was born in 1824, and when he was only 22 he was appointed to the Chair of Natural Philosophy at Glasgow, where he remained throughout his career. Skyrme quoted approvingly from a lecture Thomson (by then Lord Kelvin) gave at Johns Hopkins University in 1884, referring to James Clerk Maxwell's electromagnetic theory [12]:

"I can never satisfy myself until I can make a mechanical model of a thing. If I can make a mechanical model all the way through I can understand it. As long as I cannot make a mechanical model all the way through I cannot understand; and that is why I cannot get the electromagnetic theory."

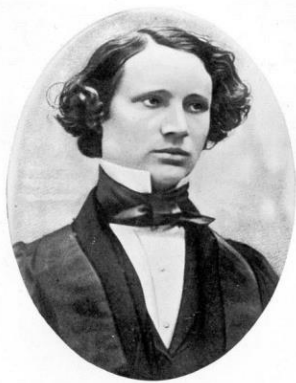
William Thomson aged 22

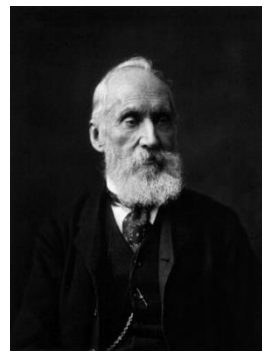
Lord Kelvin in later life

I take a slight detour here, deviating from Tony's remarks. It's actually rather surprising that Kelvin said that about Maxwell's theory, which had been published in the years 1861-1865. At an intermediate stage on the way to his electromagnetic field equations of 1865, Maxwell published a series of papers in 1861 to 1862 entitled "On Physical Lines of Force", Parts I, II and III. The second, "The Theory of Molecular Vortices Applied to the Electric Currents" [13] contained this famous diagram, showing how the compatibility of the motion of neighbouring vortices in the aether medium required the vortices to be separated by "idler wheels", whose motion generated what we now call the displacement current. As we'll soon see, Thomson was fascinated by vortices, which have more than a little in common with Skyrmions.

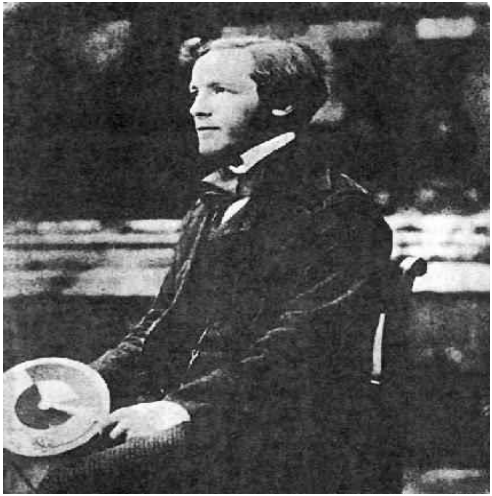 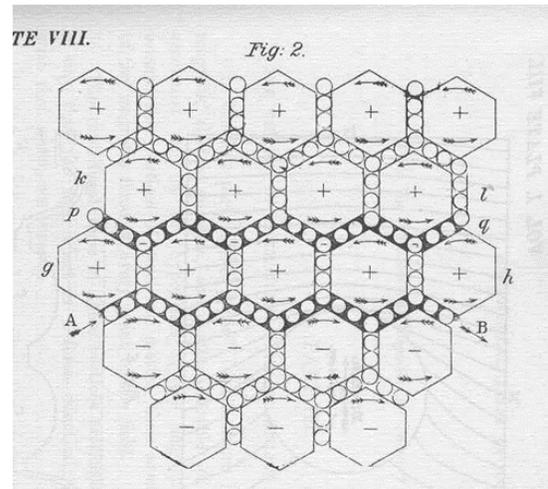

James Clerk Maxwell and his vortex aether

In fact, Tony had a direct family link with Thomson. His great grandfather on his mother's side, Edward Roberts, had been responsible in 1872-73 for the construction of the first mechanical Tidal Predictor designed by Thomson.

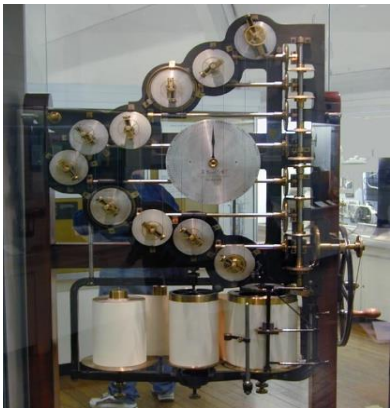 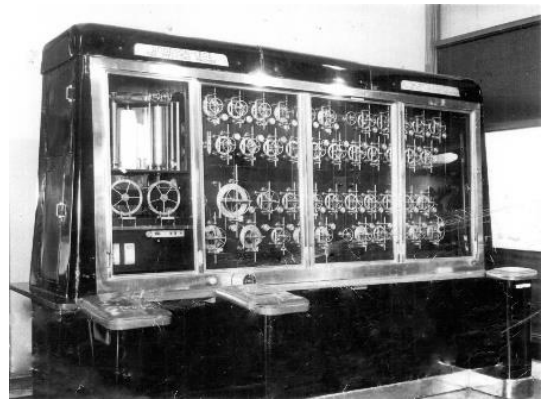

Thomson's Tidal Predictor                          Thomson-Doodson-Lege Machine c.1950

This machine stood in Tony's grandfather's house, and he told his Cosener's House audience how "the ingenuity of its mechanism, whereby it could produce this complicated pattern of tides, had considerable influence on me." This machine is now in the Science Museum in London.  A more recent one (c. 1990) is on view in the National Oceanographic Centre, Liverpool. Such machines, by the way, were of vital importance to the British Admiralty in World War Two in predicting the tides for the D-Day landings. The German Admiralty also used them: a very elaborate version can be seen in the Deutches Museum, Hamburg

Let's return to Thomson's love affair with vortices. Starting in 1867, and working with Peter Guthrie Tait, Thomson developed his "smoke ring" or "vortex" models of the atom, based on the work of Hermann von Helmholtz. The feature which so excited Thomson was that Helmholtz had proved that, in a perfect fluid, the peculiar motion he called "Wirbelbewegung" (or vortex movement, what we would now call circulation) was a constant of the motion, once created.

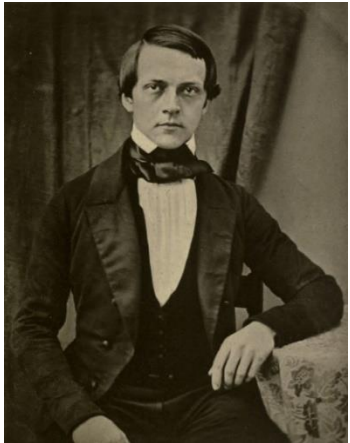 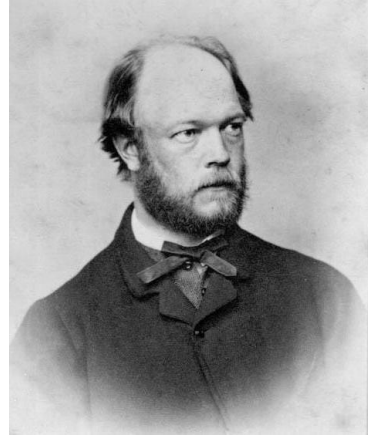

Hermann Von Helmholtz                    Peter Guthrie Tait

In 1867 Thomson read his paper "On Vortex Atoms" [14] to the Royal Society of Edinburgh. The paper begins by asserting that "the only pretext seeming to justify the monstrous assumption of infinitely strong and infinitely rigid pieces of matter [what we would call "point particles"] ….. is that urged by Lucretius and Newton – that it seems necessary to account for the unalterable distinguishing qualities of different kinds of matter". "But", Thomson continues, "Helmholtz has provided an absolute unalterable quality in the motion of any portion of a perfect fluid in which the peculiar motion which he calls "Wirbelbewegung" has been once created." So we have a spatially extended – non point-like – object carrying a conserved quality: just the kind of set-up that Tony liked. Thomson and Tait spent many happy hours experimenting with a smoke-producing machine which Tait had constructed [15]. They discovered that smoke rings could bounce off each other, and wobble, but not lose their essential integrity.

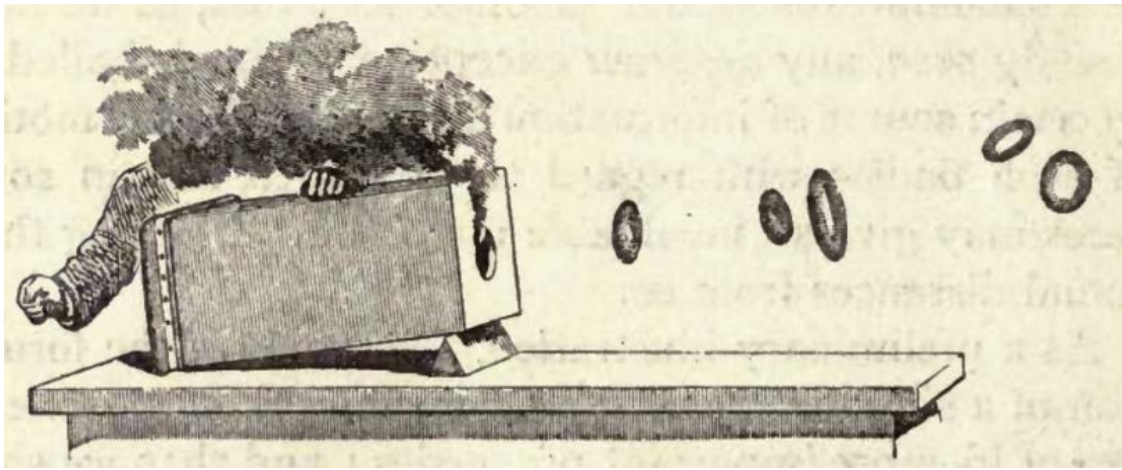

Tait's smoke machine

Thomson was led to the proposal that "all bodies are composed of vortex atoms in a perfect homogeneous liquid."

Tait became deeply interested in the mathematical classification of the various types of rings – or knots – that could be envisaged. In fact, though the vortex theory of atoms died with the discovery of the electron, Tait had laid the foundations of a major piece of mathematics: Knot Theory. Here are some of his classifications:

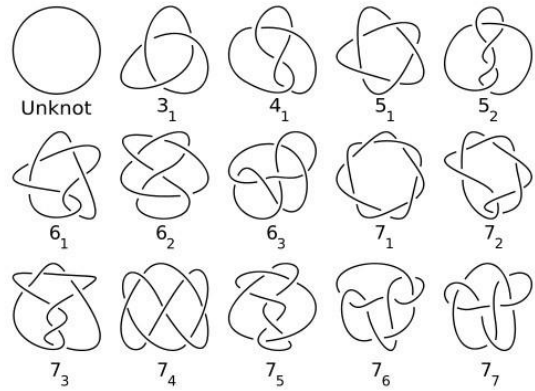

Tait's knots

And here are a few of their "vortex atoms":

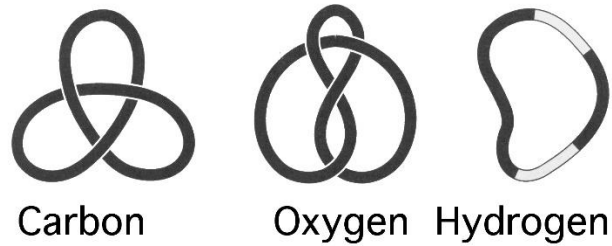

These ideas of Thomson's relate directly to the fourth of Skyrme's motivations – that of grounding a conservation law in a structural property of a classical field configuration. Another very important inspiration, which Skyrme mentioned in his talk at Cosener's House, was the Born-Infeld non-linear theory of electromagnetism. This theory [16] shares the first two of Skyrme's motivations: unification, and no point-like particles with their inevitable infinities. Born and Infeld introduce their theory by advocating a "unitarian standpoint", which "assumes only *one* physical entity, the electromagnetic field", thus following Skyrme's first motivation. In this approach, the particles of matter (such as the electron) are not considered to be entities quite distinct from the field (as in the "dualistic standpoint") but appear in the theory as singularities of the field. Actually, the singularities are present only in an idealized limit: the charge density of a particle is in fact spread out, and its electromagnetic mass is finite, satisfying Skyrme's second motivation.

So much for the historical background behind Skyrme's revolutionary ideas. In summary, he was seeking a theory in which fermionic sources of baryon number might emerge as singularities (in some limit) of a non-linear meson field theory, with baryon number corresponding to some structural property of the field configuration. The true baryon would be an extended object with finite energy.

I turn now to the second part of my talk, which will be a brief outline of how Tony implemented his program in the six papers I-VI. We need to bear in mind at the start that the enterprise was by no means mainly formal: on the contrary, the aim was to provide a practical theory of the strong interactions of mesons and baryons. The core of his ideas is contained in the first series of four papers published in the years 1958-61, which he referred to as I – IV. Most of the essential points arise already in I, entitled "A non-linear theory of strong interactions" [1]. Here Skyrme introduced four meson fields $\phi_a$, with the index a running from 1 to 4, where the first three fields are the usual isospin triplet of pions ($\pi_1$, $\pi_2$, $\pi_3$) and the fourth field σ is an isospin singlet. These four fields are subject to the constraint $\boldsymbol{\pi}^2 + \sigma^2 = f^2$ (in more modern notation), where the constant f has dimensions of mass. The fields therefore lie on the surface of a 3-sphere in four dimensions (of isospin space), leaving three degrees of freedom which are angular variables rather than linear ones. This was an essential feature of Skyrme's theory, as he explained in the third paper, III. He wrote: "The periodicity in these variables means that they are not uniquely determined (in the classical sense) by the physical state of the system. In regions of weak field the different determinations of angle generate a set of equivalent descriptions, forming separate sheets of a multiple-valued system; but when the fields become strong these sheets may cross one another, forming singularities." It was Skyrme's intuition that such singularities would correspond to fermionic sources.

Returning to paper I, these meson fields interact with nucleon fields through conventional Yukawa couplings, and with themselves via a quartic interaction. The theory was complicated and difficult to deal with, but in what proved to be a crucial step Skyrme simplified the problem by restricting space-time and isospace to two dimensions instead of four. The meson fields may be parametrized as $\phi_1 = f \sin \vartheta/4$, $\phi_2 = f \cos \vartheta/4$, so that just one angular variable is needed. The equation of motion for the meson sector is found to be precisely the Sine-Gordon equation

$$(\partial_t)^2 \vartheta - (\partial_x)^2 \vartheta = -m^2 \sin \vartheta$$

which reduces to the Klein-Gordon equation for small $\vartheta$. This equation has a long history, to which Skyrme made no reference. There are infinitely many ground states such that $\cos \vartheta = 1$, i.e. $\vartheta = 2N\pi$ where N is a positive or negative integer, or zero. The system is required to be in a ground state at the boundaries $x = +\infty$ and $x = -\infty$. A typical static solution then interpolates from one ground state at $x = -\infty$ to another at $x = +\infty$, and is a kink-type soliton having the shape of a smoothed-out step-function, centered on some point $x_0$.

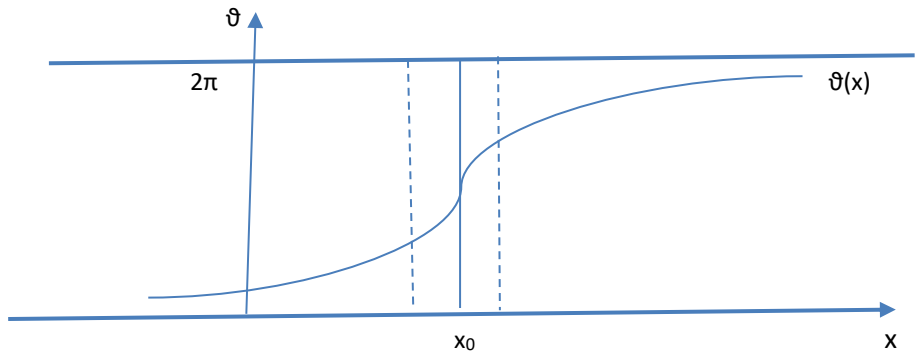

A static solution of the Sine-Gordon equation with N=1

Skyrme now identified a solution with N=1 as "a particle at $x_0$" and a solution with N=-1 as "an anti-particle at $x_0$", noting that such solutions represented localized disturbances in the field, capable of being boosted to move with arbitrary uniform velocity. He observed also that the quantity

$$\vartheta(+\infty) - \vartheta(-\infty) = (N_+ - N_-)2\pi$$

would be a constant of the motion, where $N_+$ is the number of solitons with N=+1 and $N_-$ the number with N=-1: that is, a conservation law for the number of particles minus the number of anti-particles, exactly as in the conservation of baryon number, and other similar conservation laws in fermionic field theories.

But could the kink really be a fermion? A conventional fermionic field $\psi(x)$ is a local point-like field, which would have to correspond to an idealized kink, in the form of a local discontinuous step-function singularity in the meson field – that is, the location of a jump from one "sheet" of the multiple-valued system to another "sheet". Such a correspondence, of course, could only be made in the quantized theory, which presented difficult problems. Skyrme was unable, in this first paper in the series, to prove the correspondence formally, but he gave suggestive arguments to support his fundamental idea: that all the physical features described by a Lagrangian with interacting fermion and meson fields, when treated perturbatively, could be equivalently described by its mesonic part alone, when non-perturbative solutions (i.e. those with "large" values of $\vartheta$) were included.

Skyrme returned to the Sine-Gordon equation in the fourth paper of the series. There he gave an explicit form, in terms of the quantized meson field operators, for a two-component field which creates a step-function soliton, which "almost" obeys anticommutation relations, and satisfies a massless Dirac equation. The essential part of such operators is the expression

$$\exp[2\pi i \int_x^\infty \chi(y)\, dy]$$

where $\chi$ is canonically conjugate to $\vartheta$. This has the effect of decreasing the value of $\vartheta$ by $2\pi$ for points to the right of the point x, thereby annihilating the step-function jump. This is reminiscent of the Jordan-Wigner transformation [17], which maps a line of Pauli spin operators onto fermionic operators. As it stands this operator does not quite satisfy the required fermionic field commutation relations, but Mandelstam [18] was subsequently able (apparently not knowing of Skyrme's work) to construct complete fermionic operators for the S-G equation, which obey the field equations of the massive Thirring model. At much the same time, Coleman [19], who acknowledged Skyrme's results, showed independently that the S-G model is equivalent to the massive Thirring model.

The Sine-Gordon model therefore provides a working realization of Skyrme's vision. But of course it is not relevant to the strong interactions of mesons and baryons in four dimensions, Skyrme's original focus. In the remarkable third paper in the series [3] Skyrme considered (in its classical aspects) a non-linear meson theory in four dimensions, with four meson fields subject to the constraint $\boldsymbol{\pi}^2 + \sigma^2 = f^2$, resulting in three angle-type variables. Here for the first time in particle physics, the concept of a winding number is introduced, which is a conserved quantity measuring (in this case) the number of times three-dimensional space is mapped by the fields onto the elementary volume of angular field space. As Tony explained to his Cosener's House audience [9], the concept is easy to illustrate in the one-dimensional S-G case. Consider a soliton solution $\vartheta(x)$ with general N, so that $\vartheta(-\infty) = 0$ and $\vartheta(+\infty) = 2N\pi$. If all physical quantities depend only on $\vartheta \bmod(2\pi)$ then, given the boundary conditions that $\vartheta$ tends to a multiple of $2\pi$ – i.e. to a ground state configuration – at x = +∞ and x= -∞, the real line is effectively compactified into a circle, $S_1$. $\vartheta$ is itself defined on an $S_1$, so $\vartheta(x)$ provides a mapping from the $S_1$ of real space to the $S_1$ of field space, the number of times the $\vartheta$ circle is covered being the winding number of the mapping.

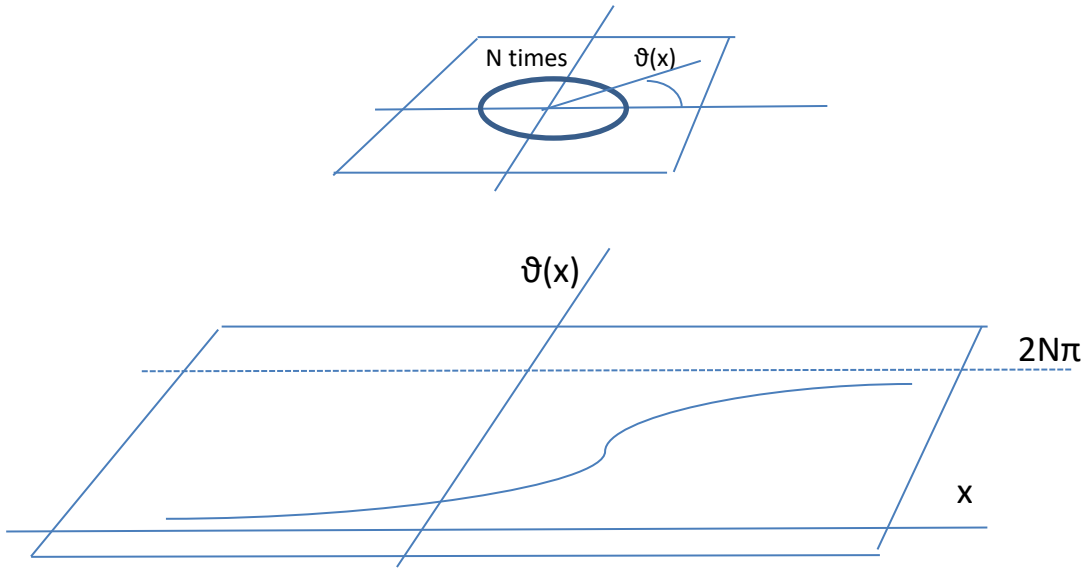

The mapping from real space to field space provided by $\vartheta(x)$

The generalization of this idea to three dimensions was one of the key contributions of this third paper. As already noted, in this case there will be three angle-type pion field variables. These variables, arranged in the configuration of a soliton, must somehow provide a mapping between real 3-D space and pion field space. What is wanted is a recipe for determining the relative orientation of the real space and isospace coordinate frames. Here Skyrme was inspired by work by Pauli [20] on the strong-coupling theory of pion-nucleon interactions. Pauli had introduced an orthogonal matrix specifying the relative orientation of the nucleon spin and the pion isospin frames. Skyrme proposed to use Pauli's orthogonal matrix to specify the orientation of the space and isospace frames in the region containing the soliton. In the simplest case, the two frames could be aligned, suggesting the ansatz

$$\pi_i = f\, r_i \sin \vartheta(r) / r\,,\quad \sigma = f \cos \vartheta(r),$$

where $\vartheta(r)$ is a smooth function of the distance r from the particle, tending to zero at large distances and to $\pi$ at the origin. This form is often called a "hedgehog" configuration, since the "quills" represented by the pion fields point outwards along the direction of the unit vector $r_i/r$. Skyrme introduced the appropriate definition of the winding number N:

$$N = -(1/2\pi^2) \int \det(\phi, \partial_x\phi, \partial_y\phi, \partial_z\phi)\, d^3\mathbf{x}$$

and showed that such a configuration carried the value N=1.

The dynamics of this classical non-linear theory raised a new question. Skyrme noted that the Lagrangian should contain a term of the form

$$\mathcal{L}_1 = \tfrac{1}{2}\, \partial_\mu \phi_a\, \partial^\mu \phi_a$$

where μ runs over x, y, z and t, and $\phi_a = (\pi_1, \pi_2, \pi_3, \sigma)$. But a simple scaling argument shows that any soliton-like solution will collapse to zero size and zero energy (an example of a result due to Derrick [21]). Skyrme therefore added a term of fourth order in the derivatives of ϕ, for which the energy will scale as the inverse size of the soliton, thereby giving stability. This "Skyrme term" has the form

$$\mathcal{L}_{SK} = (\varepsilon/f^4)\, [(\partial_\mu \phi_a\, \partial_\nu \phi_a)(\partial^\mu \phi_b\, \partial^\nu \phi_b) - (\partial_\mu \phi_a\, \partial^\mu \phi_a)^2]$$

where ε is a dimensionless constant. The theory thus employs two parameters, the energy scale parameter f and the dimensionless ε.

Skyrme then obtained upper and lower bounds on the soliton mass, in terms of these two parameters. He went on to consider adiabatic rotation of the soliton, showing that the rotational state would have equal values of the total angular momentum and isospin, as in the known Δ (3,3) resonance in pion-nucleon scattering (this is essentially a consequence of the hedgehog ansatz). Skyrme ended this classic paper by giving arguments to suggest that, as in the S-G case, the idealized point-like soliton would obey a 4-dimensional Dirac equation.

The sixth paper in the series [6] was published in 1962, and carried further the phenomenology of the soliton model of the baryon proposed in III. The equation of motion following from minimizing the energy functional was solved numerically, with the assistance of Tony Leggett, then a vacation student at Harwell. The results showed a number of encouraging resemblances between the consequences of the model and the phenomena of pion-nucleon interactions, but the difficult problem of fully quantizing the theory remained to be solved. In particular, Tony was unable to construct operators which created or destroyed the sources represented by the solitons in the point-like limit, as had been done for the S-G case. That remains an unsolved problem.

At the time Skyrme's six papers were appearing, theorists working on strong interactions had mostly given up on quantum field theory, and had turned instead to S-matrix theory and dispersion relations, and to group-theoretic classification schemes. His revolutionary ideas, couched in the then unfashionable Lagrangian framework, went largely unnoticed. In the decade following the publication of the series of six papers, his work was rarely cited: paper III, for instance, received some ten citations. Later in the decade, quarks were established as the constituents of hadrons, and field theory once again became the dominant paradigm. But strong interactions were now those between quarks and gluons, as described by quantum chromodynamics (QCD), rather than the (less fundamental) interactions of pions and nucleons. There was a modest increase of interest in Skyrme's work in the following decade, as various types of field-theoretic solitons were discovered – the citations rose to about ten per year. Then in 1983 that number increased by a factor of ten, where it has remained ever since. Google Scholar now gives over 3,000 citations for paper III.

The event triggering the explosion of interest in Tony's work was the publication of two papers by Ed Witten in 1983, entitled "Global Aspects of Current Algebra" [7] and "Current Algebra, Baryons and Quark Confinement" [8]. These papers took up earlier results due to 't Hooft [22] and Witten [23], working independently. These authors considered theories of QCD type with a general number of colors $N_c$, in the limit as $N_c$ tends to infinity with $g^2 N_c$ fixed, where g is the gauge coupling constant. They argued that, assuming confinement of particles carrying the color charges of QCD, theories of this type are equivalent to theories of mesons and glueballs alone, in which the quartic meson-meson couplings are proportional to $1/N_c$. The theory would then be weakly interacting in the large $N_c$ limit. But where would the baryons be in such an equivalent theory? Witten's answer in his paper of 1979 [23] was that they must somehow be related to solitons in the bosonic theory. The crucial motivation for this suggestion was that 't Hooft and Witten had noted that the mass of a baryon would be proportional to $N_c$ in the large $N_c$ limit. This is inversely proportional to the meson-meson coupling strength – a behavior associated with soliton states in general. When Witten made this proposal in 1979 he was evidently unaware of Skyrme's work, and was unable to identify the relevant soliton. He made the vital connection in his papers of 1983, giving new and compelling arguments for identifying Tony's solitons with the baryons of QCD in the large $N_c$ limit.

By the time of Witten's papers, the first term in Skyrme's model Lagrangian, $\mathcal{L}_1$, had an established place in the context of pion physics. There it was known as the non-linear σ-model, the σ field being eliminated via the constraint σ=√(f$^2$ - **π$^2$** ). Pions were understood to be the (approximately) massless Goldstone bosons of a spontaneously broken chiral symmetry possessed by QCD in the limit of massless u and d quarks. The Lagrangian $\mathcal{L}_1$, with f interpreted as the pion decay constant, was known to generate correctly the amplitudes for all multi-pion scattering processes (for example pion-pion scattering), up to order $p^2$ in momenta. In this context, $\mathcal{L}_1$ was to be used at tree level (no loops), so that it was effectively a classical field Lagrangian, consistent with loop corrections being suppressed in the large $N_c$ limit. This is exactly how Skyrme wanted to use it, for soliton physics. Thus the remarkable prospect emerged: the same Lagrangian which describes low energy pion scattering also describes, via its soliton solutions, the nucleons.

It would not be appropriate here to continue the story of Tony's solitonic nucleons in any further detail. It may be fair to say that, despite yielding a new and insightful way of understanding how QCD produces nucleons, the Skyrme Lagrangian cannot offer a systematic quantitative approach to the calculation of nucleon properties. Difficulties include the fact that the fourth-order "Skyrme term" lacks any fundamental connection to QCD, and must be regarded as purely phenomenological. Indeed, it is likely that the bosonic theory envisaged by 't Hooft and Witten involves infinitely many meson fields. Nevertheless, Tony's model stands as a remarkable *effective* field theory, capable of describing – with only one free parameter - a wide range of pion-nucleon physics with, at worst, 30% accuracy (recall that $N_c$=3!).

But of course this very meeting is itself a tribute to the fundamental nature of Skyrme's ideas: like all profound contributions to physics, his introduction of a new type of conservation law, associated with a winding number – or as we now term it, a topological quantum number – has proved to have applications far beyond his original theory of mesons and baryons – in particular to two-dimensional physics, a case Tony never himself considered, as far as I know. And his studies of the Sine-Gordon-Thirring correspondence provided the first example of such remarkable equivalences, which continue to reveal the richness of quantum field theory.

### Acknowledgments

I thank Professors Peter Hatton and Christian Pflederer for their kind invitation to present this talk at the WE-Heraeus-Seminar on Skyrmions in magnetic materials. I am very grateful to the Wilhelm and Else Heraeus Foundation for its generous support.

### References

1. T. H. R. Skyrme, *A non-linear theory of strong interactions*, Proc. Roy. Soc. A **247**, 260-278 (1958).

2. T. H. R. Skyrme, *A unified model of K- and π-mesons*, Proc. Roy. Soc. A **252**, 236-245 (1959).

3. T. H. R. Skyrme, *A non-linear field theory*, Proc. Roy. Soc. A **260**, 127-138 (1961).

4. T. H. R. Skyrme, *Particle states of a quantized meson field*, Proc. Roy. Soc. A **262**, 237-245 (1961).

5. T. H. R. Skyrme and K. J. Perring, *A model unified field equation*, Nucl. Phys. **31**, 550- 555 (1962).

6. T. H. R. Skyrme, *A unified field theory of mesons and baryons*, Nucl. Phys. **31**, 556-569 (1962).

7. E. Witten, *Global aspects of current algebra*, Nucl. Phys. **B 223**, 422- 432 (1983).

8. E. Witten, *Current algebra, baryons and quark confinement*, Nucl. Phys. **B 223**, 433- 444 (1983).

9. T. H. R. Skyrme, *The Origins of Skyrmions*, in *A Breadth of Physics*, Proc. Peierls 80[th] Birthday Symposium, Eds. R. H. Dalitz and R. B. Stinchcombe, World Scientific Press, Singapore (1988), pages 193-202.

10. *ibid.* pages 205-219.

11. R. H. Dalitz, *An outline of the life and work of Tony Hilton Royle Skyrme (1922-1987)*, Int. J. Mod. Phys. **A 3**, 2719-2744 (1988).

12.*.* Lord Kelvin (Sir William Thomson), *Notes of Lectures on Molecular Dynamics and the Wave Theory of Light delivered at the Johns Hopkins University, Baltimore, by Sir William Thomson*, stenographic report by A. S. Hathaway, (Johns Hopkins University, Baltimore, 1884), pages 835-836.

13. J. C. Maxwell, *On physical lines of force. Part II: The theory of molecular vortices applied to electric currents*, Phil. Mag. **21**, 281-291 (1861).

14. Lord Kelvin (Sir William Thomson), *On Vortex Atoms*, Proc. Roy. Soc. Edinb. **VI**, 94-105 (1867).

15. P. G. Tait, *Lectures on Some Recent Advances in Physical Science*, MacMillan & Co., London, 2[nd] edtn. 1876.

16. M. Born and L. Infeld, *Foundations of the New Field Theory*, Proc. Roy. S oc. **144**, 425-451 (1934).

17. P. Jordan and E. P. Wigner, *About the Pauli exclusion principle* (in German), Z. Phys. **47**, 631-651 (1928).

18. S. Mandelstam, *Soliton operators for the quantized Sine-Gordon equation*, Phys. Rev. **D 11**, 3026-3030 (1975).

19. S. Coleman, *The Quantum Sine-Gordon equation as the massive Thirring model*, Phys. Rev. **D 11**, 2088-2097 (1975).

20. W. Pauli, *Meson Theory of Nuclear Forces*, Interscience, New York, 1946.

21. G. H. Derrick, *Comments on nonlinear wave equations as models for elementary particles*, Math. Phys.**5**, 1252-1254 (1964).

22. G. 't Hooft, *A planar diagram theory for strong interactions*, Nucl. Phys. **B 72**, 461-473 (1974); *ibid. A two-dimensional model for mesons*, **B 75**, 461- 470 (1974).

23. E. Witten, *Baryons in the 1/N expansion*, Nucl. Phys. **B 160**, 57- 115 (1979).